\documentclass[12pt,preprint]{aastex}

\usepackage{graphicx,epsfig,natbib,color}
\usepackage{lscape}
\usepackage{graphicx}
\usepackage{epsfig}
\usepackage{natbib}
\usepackage{multirow}

\begin{document}
\title{Extremely Large EUV Late Phase of Solar Flares}

\author{Kai Liu\altaffilmark{1,2}, Yuming Wang\altaffilmark{1}, Jie Zhang\altaffilmark{3}, Xin Cheng\altaffilmark{4}, Rui Liu\altaffilmark{1}, Chenglong Shen\altaffilmark{1}}

\affil{$^1$ CAS Key Laboratory of Geospace Environment, Department of Geophysics and Planetary Sciences, School of Earth and Space Science, University of Science and Technology of China, Hefei 230026, China}\email{kailiu@mail.ustc.edu.cn}
\affil{$^2$ Key Laboratory of Solar Activity, National Astronomical Observatories, Chinese Academy of Sciences, Beijing 100012}
\affil{$^3$ School of Physics, Astronomy and Computational Sciences, George Mason University, 4400 University Drive, MSN 6A2, Fairfax, VA 22030, USA}
\affil{$^4$ School of Astronomy and Space Science, Nanjing University, Nanjing 210093, China }

\begin{abstract}

The second peak in the Fe XVI 33.5 nm line irradiance observed during solar flares by \emph{Extreme ultraviolet Variability Experiment} (EVE) is known as Extreme UltraViolet (EUV) late phase. Our previous paper \textbf{\citep{Liu2013}} found that the main emissions in the late phase are originated from large-scale loop arcades that are closely connected to but different from the post flare loops (PFLs), and we also proposed that a long cooling process without additional heating could explain the late phase. In this paper, we define the extremely large late phase because it not only has a bigger peak in the warm 33.5 irradiance profile, but also releases more EUV radiative energy than the main phase. Through detailedly inspecting the EUV images from three point-of-view, it is found that, besides the later phase loop arcades, the more contribution of the extremely large late phase is from a hot structure that fails to erupt. This hot structure is identified as a flux rope, which is quickly energized by the flare reconnection and later on continuously produces the thermal energy during the gradual phase. Together with the late-phase loop arcades, the fail to erupt flux rope with the additional heating create the extremely large EUV late phase.

\end{abstract}


\section{Introduction}

Solar flare is one of the most energetic phenomena in the solar atmosphere, manifested as sudden and rapid enhancement of intensity in almost all electromagnetic wavelengths. It is widely accepted that the fast magnetic reconnection in a stretched current sheet (CS) is the physical mechanism that releases the stored magnetic energy in the corona and subsequently produces the observed flare. Flares, especially the strong ones, are often accompanied with a coronal mass ejection (CME), which are called eruptive flares. On the other hand, the ones without an accompanying CME are called confined flares \citep{Moore2001,Wang2007}.

The magnitude of a solar flare is defined by its peak soft X-ray (SXR) flux, according to the 1 - 8 {\AA} observations made by \emph{Geosynchronous Operational Environment Satellite X-Ray Sensor} (GOES/XRS). In temporal evolution, a flare typically consists of two phases, the impulsive phase and gradual phase \citep[e.g.][]{Benz2008,Hudson2011}. The impulsive phase is characterized by the intense increase of electromagnetic emissions lasting from a few minutes to tens of minutes, in particular in non-thermal signatures such as hard X-rays; this impulse of energy release usually corresponds to the rise phase of the flare's soft X-ray profile, as related to the Neupert Effect \citep{Neupert1968,Kane1980}. Following the impulsive phase, the soft X-ray emission tends to gradually decrease and return to its original level, thus the gradual phase, which is also often called the decay phase. Statistically, flares with SXR duration greater than 2 hours, the so-called long-duration-events (LDEs), have a higher possibility to be eruptive \citep{Harrison1995,Yashiro2009}.

Recently, \citet{Woods2011} found a new phase of solar flares using the high resolution extreme ultraviolet (EUV) irradiance observations from the \emph{Extreme ultraviolet Variability Experiment} (EVE; \citealt{Woods2012}) onboard the \emph{Solar Dynamics Observatory} (SDO; \citealt{Pesnell2012}). They found that certain flares had two distinct flux peaks in the EVE warm line 33.5 nm (Fe XVI; $\sim$2.5 MK) irradiance profile: the first peak nearly coincides with the GOES SXR peak, while the second peak is about tens of minutes to a few hours later than the first peak. While the first peak occurs in the flare impulsive phase or main phase, the second peak is coined as the EUV late phase. \citet{Woods2011} also presented the flux ratio of the second peak to the first peak, which ranges from 0.2 to 4.1, with the average value of 0.8.

By combining the EVE irradiance observations with the spatially-resolved EUV image observations from the \emph{Atmospheric Imaging Assembly} (AIA; \citealt{Lemen2011}) also onboard SDO, \citet{Liu2013} found that the emission during the EUV late phase originates from a separate set of large loop arcades other than the source of the main flare that produces a smaller-sized post-flare loops (PFLs). These large loop arcades, also called late phase loop arcades, were found to be magnetically connected with the PFLs. \citet{Liu2013} also suggested that, even though the heating might occur at the same time in the large arcade as in the smaller loops responsible for the main phase, the longer plasma cooling process in the large loop arcade be responsible for the creation of the EUV late phase. Other works based on model calculations pointed out that there might be additional heating during the gradual phase when a EUV late phase occurred \citep[e.g.][]{Hock2012,Sun2013,Li2014}.

In this paper, we present a case study of the solar flare on November 05, 2010, which produced an extremely large EUV late phase. This extremely large EUV phase has a stronger second peak in EUV (ratio of $\sim$ 2.1 between the second and the first peak). The calculation indicated that the late phase contains, much more EUV radiative energy than in the main phase (ratio $\sim$ 3.4). We further identified that the extremely large EUV phase originated from two sources: a stationary large loop arcade (as in a usual late phase flare) and an erupted-but-failed hot structure. The failed structure is presumably a magnetic flux rope. We believe that it is the hot structure produced by the failed flux rope eruption that created the extremely large EUV late phase, thanks to the continuous energy release in the trapped magnetic flux rope. Observations and results are presented in Section 2, Discussions are given in Section 3, and Section 4 is for Conclusions.

\section{Observations and Results}

The event we studied in this paper is an M1.0 solar flare occurred on Nov. 5th, 2010. Based on the GOES SXR profile (red curve in Figure~\ref{f1}), this flare started at 12:43 UT, peaked at 13:29 UT (red vertical solid line a), and followed by a long decay phase lasting about 4 hours; therefore, it is a LDE flare event. However, through carefully checking the coronagraph images from \emph{Large Angle and Spectrometric Coronagraph C2} (LASCO/C2; \citealt{Howard1992}) onboard \emph{Solar and Heliospheric Observatory} (SOHO; \citealt{Domingo1995}), we found no CME associated with this flare (\textbf{see the online LASCO animation} and Figure~\ref{f4}a). It is unusual that such a long duration flare is a confined event. We shall explore the connection of this property with the late phase flare.

\subsection{Extremely Large EUV Late Phase}

EVE observations measures the solar EUV differential irradiance from 0.1 nm to 105 nm with unprecedented spectral resolution of 0.1 nm, temporal cadence of 10 s, and accuracy of 20\% \citep{Woods2012}. The EVE Level-2 processing produces a combined set of merged spectra provided in a pair of files: one file (EVS) contains the full spectra, and the other (EVL) contains the isolated lines from ionized solar elements and the bands simulated for other instruments such as AIA.

\begin{figure} 
     \vspace{-0.0\textwidth}    
     \centerline{\hspace*{0.02\textwidth}
               \includegraphics[width=0.9\textwidth,clip=1]{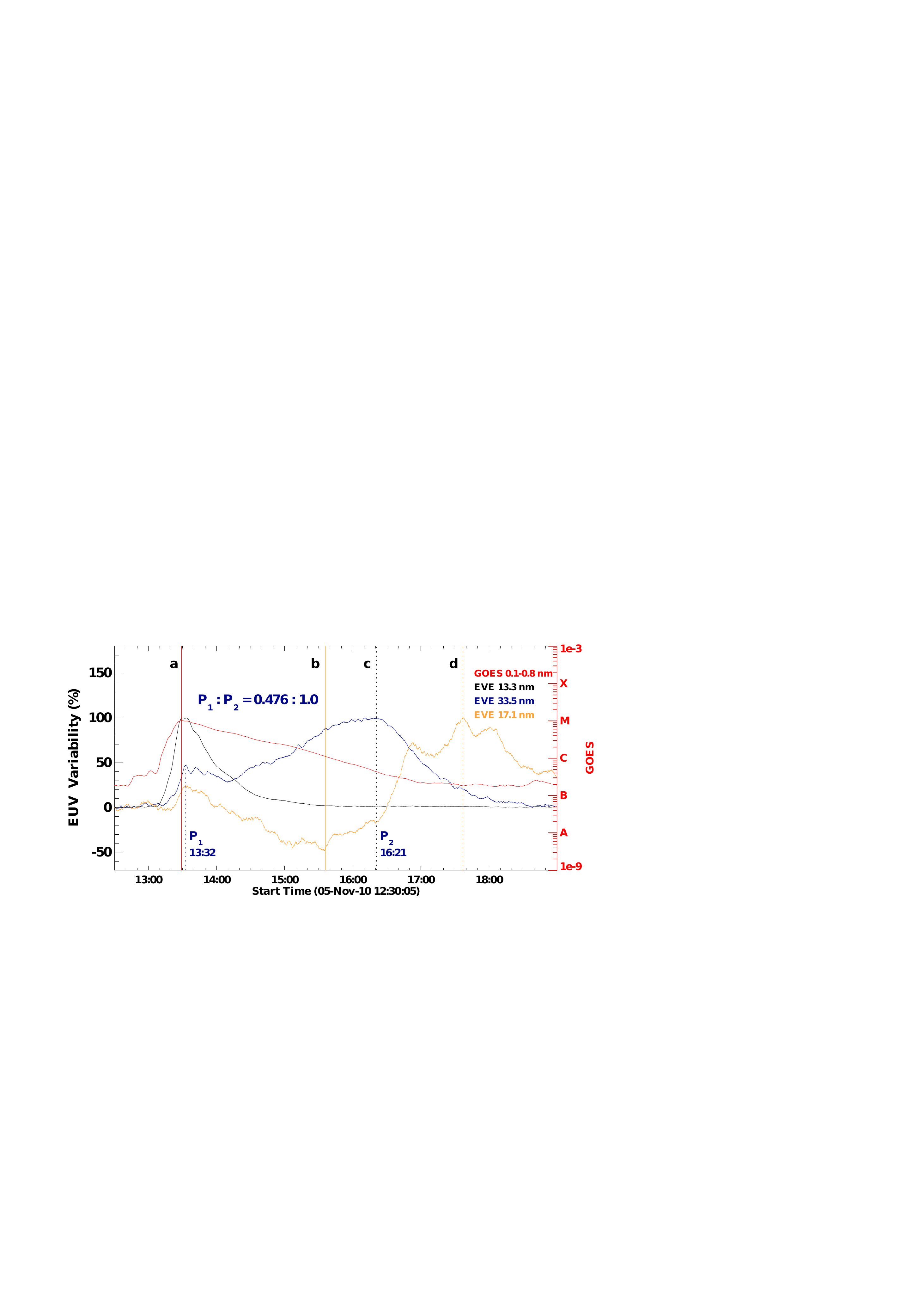}
               }

\vspace{0.0\textwidth}   
\caption{The temporal variation of the differential irradiance (normalized to the peak intensity) in three different spectral lines of SDO EVE observation of the M1.0 flare on 2010 November 5. The GOES SXR 0.1 - 0.8 nm profile is also shown (red curve). Vertical lines a, b, c and d indicate the flare's main phase peak (13:29 UT), EVE 171 {\AA} dimming valley (15:36 UT), EVE 335 {\AA} late phase peak (16:21 UT), and EVE 171 {\AA} late phase peak (17:37 UT), respectively.}
\label{f1}

\end{figure}

We used EVL data to study the EUV late phase of this flare event. We chose three lines with their central wavelength of 13.3 nm (Fe XX, $\sim$10 MK), 33.5 nm (Fe XVI, $\sim$2.5 MK) and 17.1 nm (Fe IX, $\sim$0.7 MK), corresponding to high temperature, warm temperature and cool temperature, and plot their irradiance variabily (normalized) as black, blue and yellow curves in Figure~\ref{f1}, respectively. Similar to GOES SXR profile, the profile of high temperature line 13.3 nm only had a single peak, and peaked at almost the same time ($\sim$13:29 UT) as in X-ray. However, in the profile of warm 33.5 nm line, there was a distinct second peak (P$_{2}\sim$16:21 UT, blue vertical dotted line c in Figure~\ref{f1}) other than the peak in the main phase (P$_{1}\sim$13:32 UT); the second peak lagged about 170 minutes behind the first peak. The existence of this second peak confirmed that this flare did have a EUV late phase. Moreover, the peak value of the late phase is much larger than that in the main phase; the ratio between the second and the first for peaks is as high as 2.1.

The profile of the cool line 17.1 nm had two peaks: the main phase peak at $\sim$13:32 UT and the late phase peak at $\sim$17:37 UT, yellow vertical dotted line d in Figure~\ref{f1}. We note that, during the late phase, there were multiple peaks, and we chose the largest one representing the peak time. Obviously, there was a strong dimming in the cool line between the two phases. Immediately following the main phase peak, the irradiance of the cool 17.1 nm line began to decrease until it reached its minimum around 15:36 UT. The flux intensity at the minimum was significantly lower than the pre-flare background level. The flux then had a slow recovery phase until 16:21 UT, which was followed by a faster ascending phase. Nothe that, the transition time between the slow recovery to fast ascending coincides well with the peak time of the late phase in 33.5 nm line. Based on the fact that this is a confined flare, it could be concluded that the dimming or depression in the cool line was mainly due to the thermal effect rather than the mass-loss effect.

\begin{figure} 
     \vspace{-0.0\textwidth}    
     \centerline{\hspace*{0.02\textwidth}
               \includegraphics[width=0.88\textwidth,clip=1]{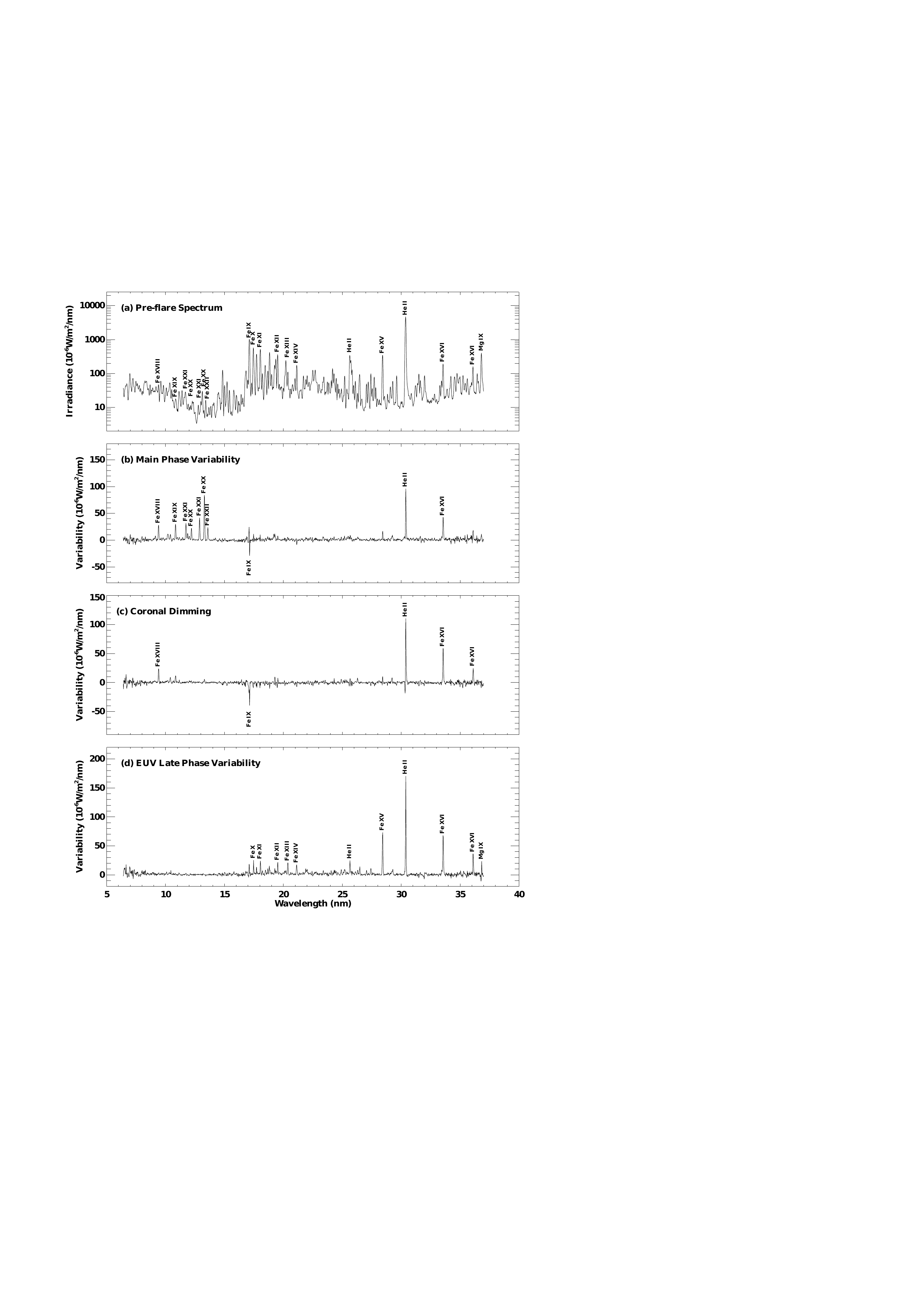}
               }

\vspace{0.0\textwidth}   
\caption{Flare spectral variations from the EVE MEGS-A channel (6-37 nm) for the M1.0 flare on 2010 Nov. 5th. Panel (a) shows the pre-flare spectrum. Panel (b)-(d) show the variability between the pre-flare (12:30 UT) irradiance and the main phase (13:29 UT; vertical line a in Figure~\ref{f1}), coronal dimming (15:36 UT; vertical line b in Figure~\ref{f1}), and EUV late phase (16:21 UT; vertical line c in Figure~\ref{f1}), respectively.}
\label{f2}

\end{figure}

\begin{figure} 
     \vspace{-0.0\textwidth}    
     \centerline{\hspace*{0.02\textwidth}
               \includegraphics[width=0.88\textwidth,clip=1]{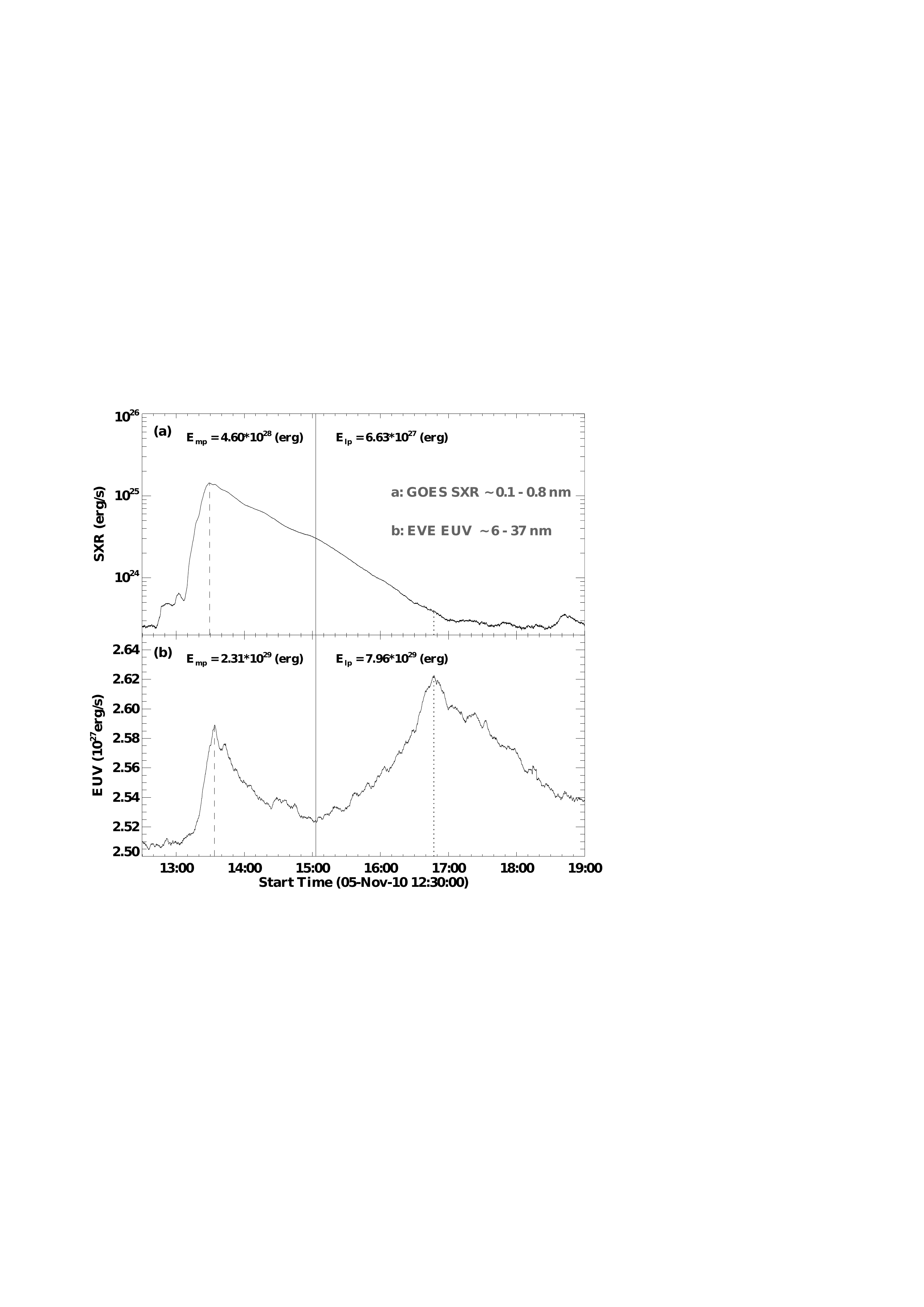}
               }

\vspace{0.0\textwidth}   
\caption{Radiative energy loss rates of two wavebands: (a), GOES Soft X-ray 0.1 - 0.8 nm; (b), EVE EUV 6 - 37 nm. }
\label{f3}

\end{figure}

The EVS spectral data product provided the spectrum irradiance variability of the flare within the wavelength range of 6 - 37 nm (see Figure~\ref{f2}(a)). The units for EVE spectrum irradiance were $watt/m^{2}/nm$. Integrating over this wavelength range, we obtained the total EUV radiative energy loss rate (units $J/m^{2}/s$) of EVE EUV spectrum at Earth (1 AU). We further converted the radiative energy loss rate to energy loss at the Sun ($ergs/s$) by multiplying a factor of $1.406\times10^{30}$~($=10^{7}\cdot2\cdot\pi\cdot(1 AU)^{2}$), assuming a uniform angular distribution of the radiation \citep{Woods2006}. The same conversion method has been applied to GOES SXR observations, and the results are displayed in Figure~\ref{f3}.

It was interesting to note that there were also two peaks in the time series of the radiative energy loss rates of EVE EUV spectrum, while only a single peak in the GOES SXR profile. The time of the first peak is 13:34 UT, lagging about 5 minutes behind the GOES X-ray peak (13:29 UT). This short lag is likely the consequence of the PFLs' cooling effect. The  time of the second peak time is 16:47 UT, later than the 33.5 nm late phase peak and earlier than the 17.1 nm's. Apparantly, the second peak is larger than the first peak in energetics. From the energy release point of view, we defined these two peaks as the peaks of the main phase and the EUV late phase. The separating time between the two phases is at about 15:03 UT (see Figure~\ref{f3}, vertical solid line). For calculating the flare energy, we subtracted the pre-flare value from the energy loss rates (we use 12:30 UT as the pre-flare time hereinafter), then integrated them in the time range of both phases. As shown in Figure~\ref{f3}, the total radiative energy in GOES SXR 0.1 - 0.8 nm was about $5.26\times10^{28} erg$, and the ratio of the main phase ($4.60\times10^{28} erg$) to the late phase ($6.63\times10^{27} erg$) is about 7 to 1. However in the EVE EUV 6 - 37 nm range, the total flare radiative energy is about $1.03\times10^{30} erg$, and the ratio between the main phase ($2.31\times10^{29}$) and the late phase ($7.96\times10^{29}$) is about 1 to 3.4. Obviously, the late phase had much more energy released than that in the main phase. It is thus called the event of extremely large EUV late phase.

By subtracting the spectrum irradiance at the pre-flare time (we use 12:30 UT as the pre-flare time hereinafter), we derived the EVE EUV spectrum variabilities of different phases (panel (b), (c) and (d) in Figure~\ref{f2}; we use 5 minutes average here), which were main phase peak ($\sim$ 13:29 UT), coronal dimming ($\sim$15:36 UT) and late phase peak ($\sim$ 16:21 UT), respectively. From these variabilities profiles, we identified 19 different spectral lines with significant variations ($\bigtriangleup I \geq 20$, units: $10^{-6}W/m^{2}/nm$) and listed their properties in Table~\ref{t1} (based on the chianti atomic database, \citealt{Dere1997,Landi2013}). The Table indicates that, only the cool corona line, Fe IX 17.1 nm shows the dimming. Except the choromosphere line He II 30.4 nm and warm corona line Fe XVI 33.5 nm, which had larger enhanced radiation in all phases, other lines had enhanced radiation either in the main phase (hotter lines with shorter wavelengthes, \textless 15 nm) or in the late phase (cooler lines with longer wavelengthes, \textgreater 15 nm).

The six lines Fe XX 13.3 nm, Fe XVIII 9.4 nm, Fe XVI 33.5 nm, Fe XIV 21.1 nm, Fe XII 19.5 nm and Fe IX 17.1 nm in Table~\ref{t1}, correspond to the six coronal passbands of AIA (Fe XX 131 \AA, Fe XVIII 94 \AA, Fe XVI 335 \AA, Fe XIV 211 \AA, Fe XII 193 \AA~ and Fe IX 171 \AA). Thus, we further inspected the AIA image sequences to study the morphology and dynamic process of the flare event.

\subsection{Failed Eruption and Hot Structure}

AIA takes high-resolution images ($4096\times4096$ pixels, 0.6" pixel size and 1.5" spatial resolution) of the whole sun with a high cadence of 12 seconds. Further, the AIA instrument has ten passbands sensitive to different temperatures, thus allowing the investigation of the thermal property of flare structures. Six of the ten passbands are sensitive to coronal temperatures as mentioned above, and we only use the images from these six passbands in this paper.

From the AIA images, we found that the flare event occurred at NOAA AR 11121, centered at the heliographic coordinates 20${^\circ}$S and 75${^\circ}$E. Therefore, it waslocated at the southeast limb allowing us to better study its structure along the radial direction in lower corona. To eliminate the interference on flux calculation from other ARs, we cut a sub-region of the AIA images containing the AR 11121 and focused on these sub-images in different wavebands. By summing all the pixel values in this sub-region and plotting its variation against time (Figure~\ref{f5}(a)), we got very similar profiles as EVE observations (Figure~\ref{f1}), which confirmed that emissions for both main phase and late phase were mainly originated from this sub-region.

\begin{figure} 
     \vspace{-0.0\textwidth}    
     \centerline{\hspace*{0.02\textwidth}
               \includegraphics[width=0.88\textwidth,clip=1]{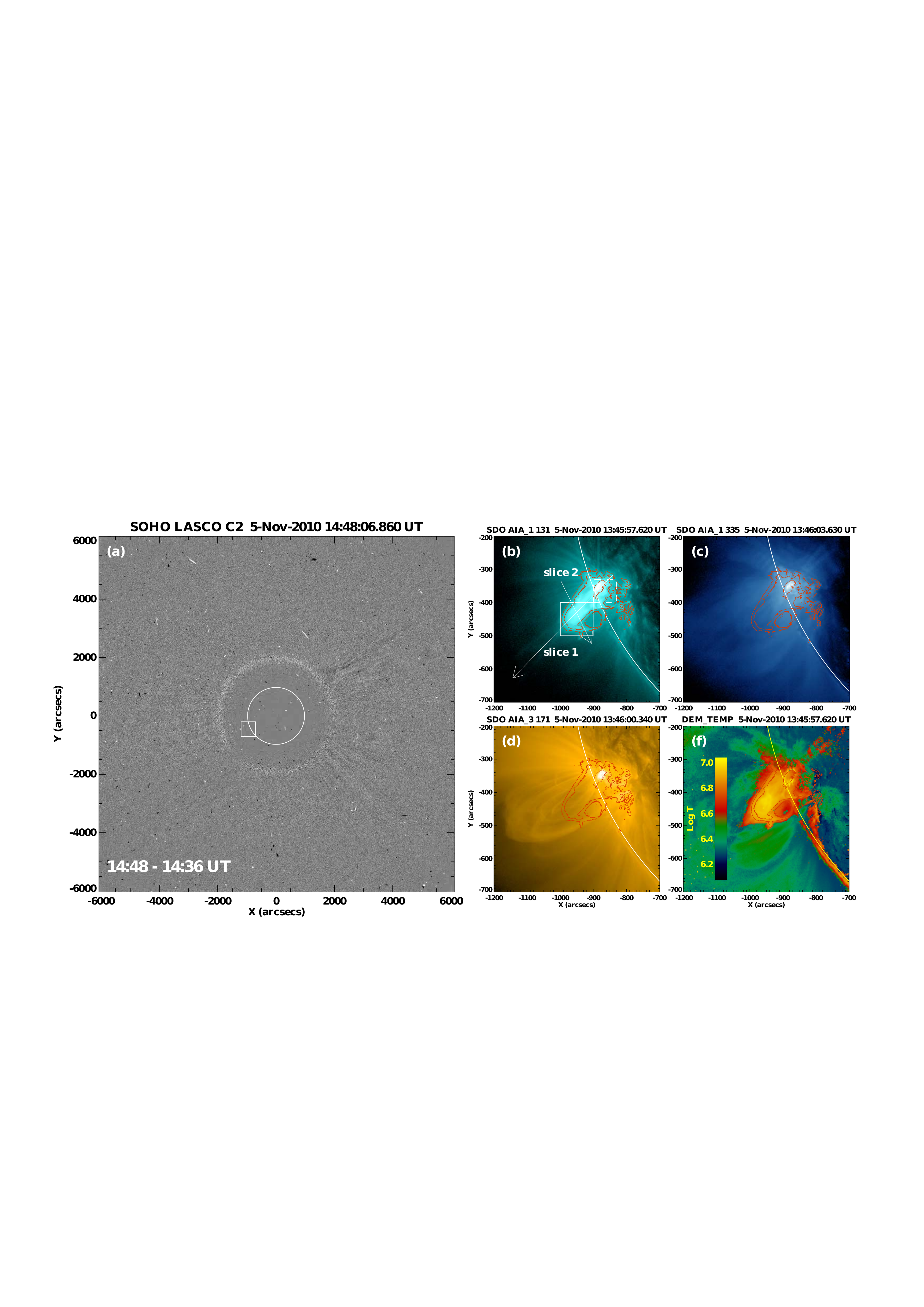}
               }

\vspace{0.0\textwidth}   
\caption{(a), SOHO/LASCO C2 difference image; white box indicates the location of AIA sub images. (b), (c) and (d) are the AIA sub images at waveband of 131 {\AA}, 335 {\AA} and 171 {\AA}, respectively; red curves are contours of bright structure in 131 {\AA} image. In panel (b), the dotted line box and solid line box indicate main flare region and late phase region for Figure~\ref{f5}, separately; two white arrows represent two slices for Figure~\ref{f6}.}
\label{f4}

\end{figure}

\begin{figure} 
     \vspace{-0.0\textwidth}    
     \centerline{\hspace*{0.02\textwidth}
               \includegraphics[width=0.90\textwidth,clip=1]{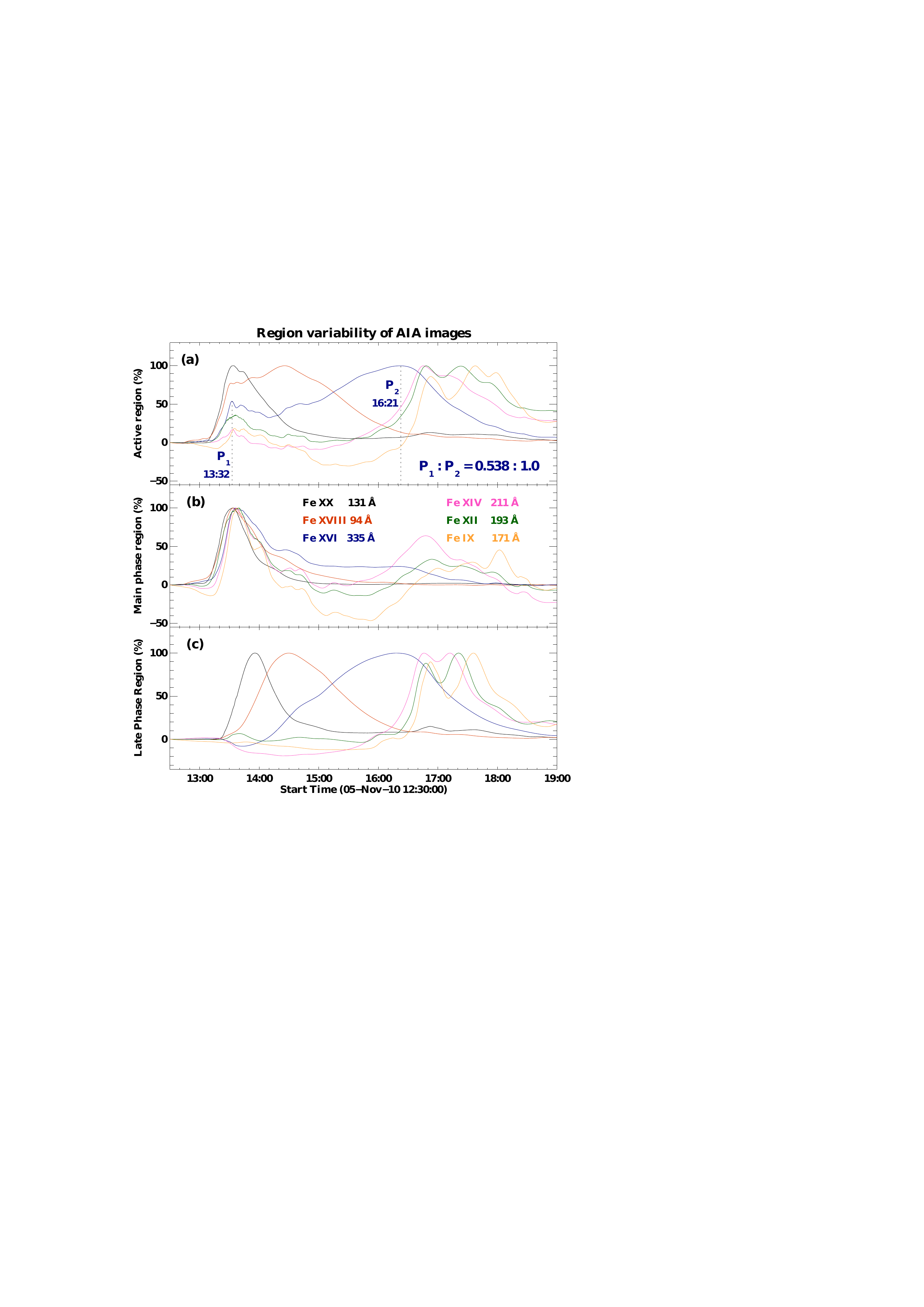}
               }

\vspace{0.0\textwidth}   
\caption{Region EUV variability of AIA sub regions. (a), the whole AIA sub image including the AR for the flare event (the white solid line box in Figure~\ref{f4}a); (b), main flare region (the white dotted line box in Figure~\ref{f4}b); (c), the late phase region (the white solid line box in Figure~\ref{f4}b). Different colors are corresponding to different passbands as shown in panel (b).}
\label{f5}

\end{figure}

An interesting high-lying hot-blob-like feature was found in these sub-region images.As the bright post-flare-loop (PFL) arcade rose and reached a certain height, the hot blob-like structure appeared above, then kept expanding and ascending until it was stopped by the `cold' overlying loops (\textbf{see the online AIA animation}). This blob-like structure was diagnosed as hot plasma (5 $\sim$ 18 MK), since it was brightening only in hot passbands, while the PFLs were bright in all passbands (see Figure~\ref{f4}(b), (c) and (d)). As the structure faded in the hot passbands, certain similar structure began to appear in the warm passband 335 \AA, coinciding with the late phase. We plotted the region variabilities of this hot structure along with the PFLs in Figure~\ref{f5}(b) and (c), corresponding to the white solid box and dotted box respectively in Figure~\ref{f4}(b). Based on these variability profiles, we confirmed that the EUV emissions for main phase were mainly from the PFLs, while the EUV late phase emissions were mainly from the area of the hot structure.

\begin{figure} 
     \vspace{-0.0\textwidth}    
     \centerline{\hspace*{0.02\textwidth}
               \includegraphics[width=1.0\textwidth,clip=1]{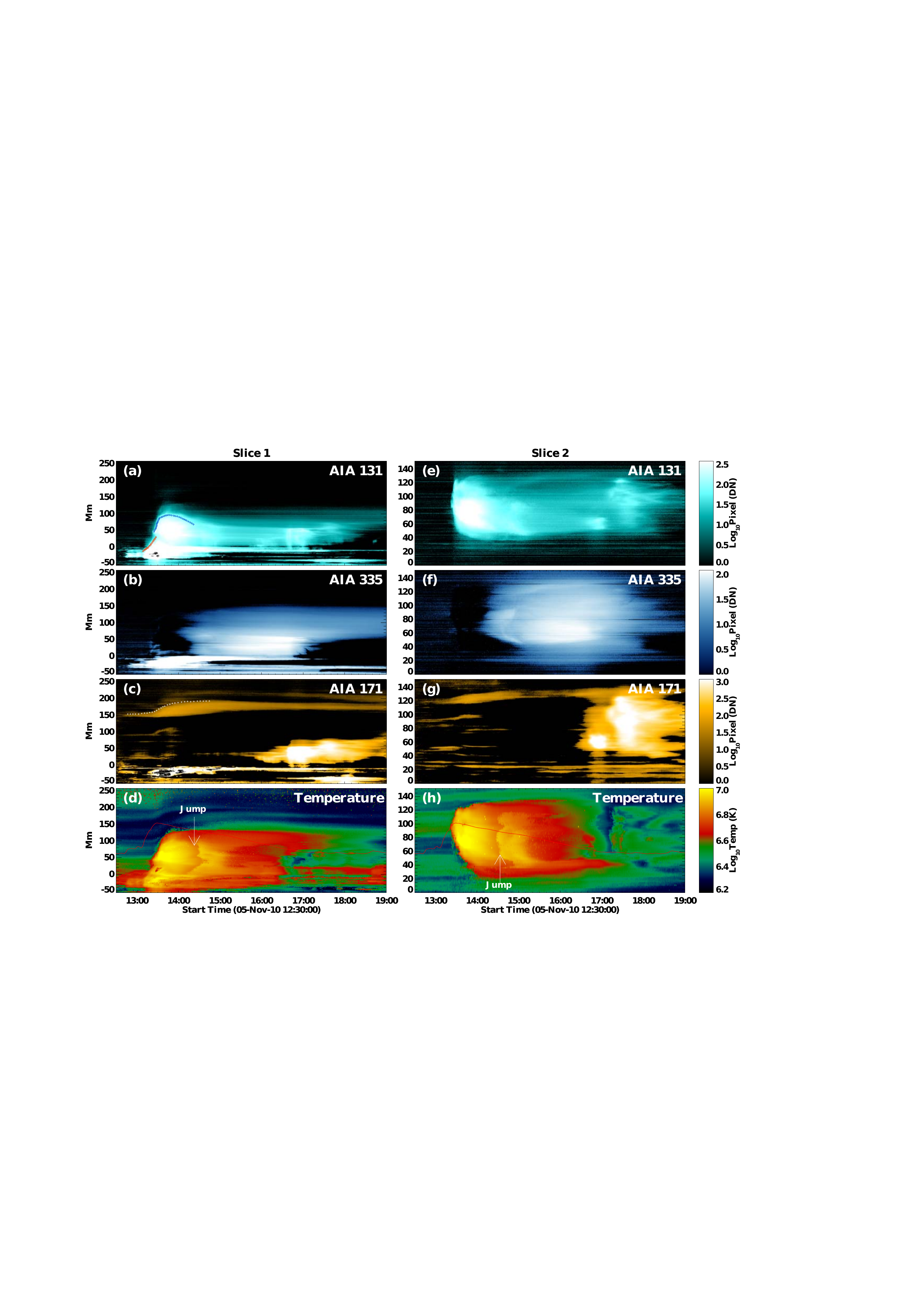}
               }

\vspace{0.0\textwidth}   
\caption{Slice-time plots of slice 1 ((a)-(d)) and slice 2 ((e)-(h)). The locations of the two slices are marked in Figure~\ref{f4}(b) as white arrows. From top to bottom, panels are corresponding to AIA 131 {\AA}, 335 {\AA} and 171 {\AA} base-difference images (the base time is 12:30 UT) and the temperature maps derived from DEM analysis results, respectively. The red triangles in panel (a) indicate the leading edges of the hot post-flare loops; the blue diamonds also in panel (a) indicate the leading edges of the hot bright structure; the white pluses in panel (c) indicate the leading edges of the cold overlying loop arcades; in both panel (d) and (h), the red curve represents the GOES SXR profile and the white arrow is pointing to a temperature jump process.}
\label{f6}

\end{figure}

To study the dynamics of this flare event, two slices were cut from the AIA difference images (hot 131 {\AA}, warm 335 {\AA} and cold 171 {\AA}) across the hot structure in two directions, the radial direction (slice 1) and transverse direction (slice 2, both slices were indicated as two arrows in Figure~\ref{f4}(a)). We stacked these slices along time as the slice-time plots in Figure~\ref{f6}. The appearance and ascending process of the hot structure was only obvious in the hot 131 {\AA} plot (Figure~\ref{f6}(a)), while the PFLs brightened in all the plots. The hot structure appeared above the PFLs and maintained a certain distance between them. The separation and sharp boundaries of the two structures allowed us to identify their evolving leading edges (red triangles for the PFLs and blue diamonds for the hot structure in Figure~\ref{f6}(a)).

\begin{figure} 
     \vspace{-0.0\textwidth}    
     \centerline{\hspace*{0.02\textwidth}
               \includegraphics[width=0.77\textwidth,clip=1]{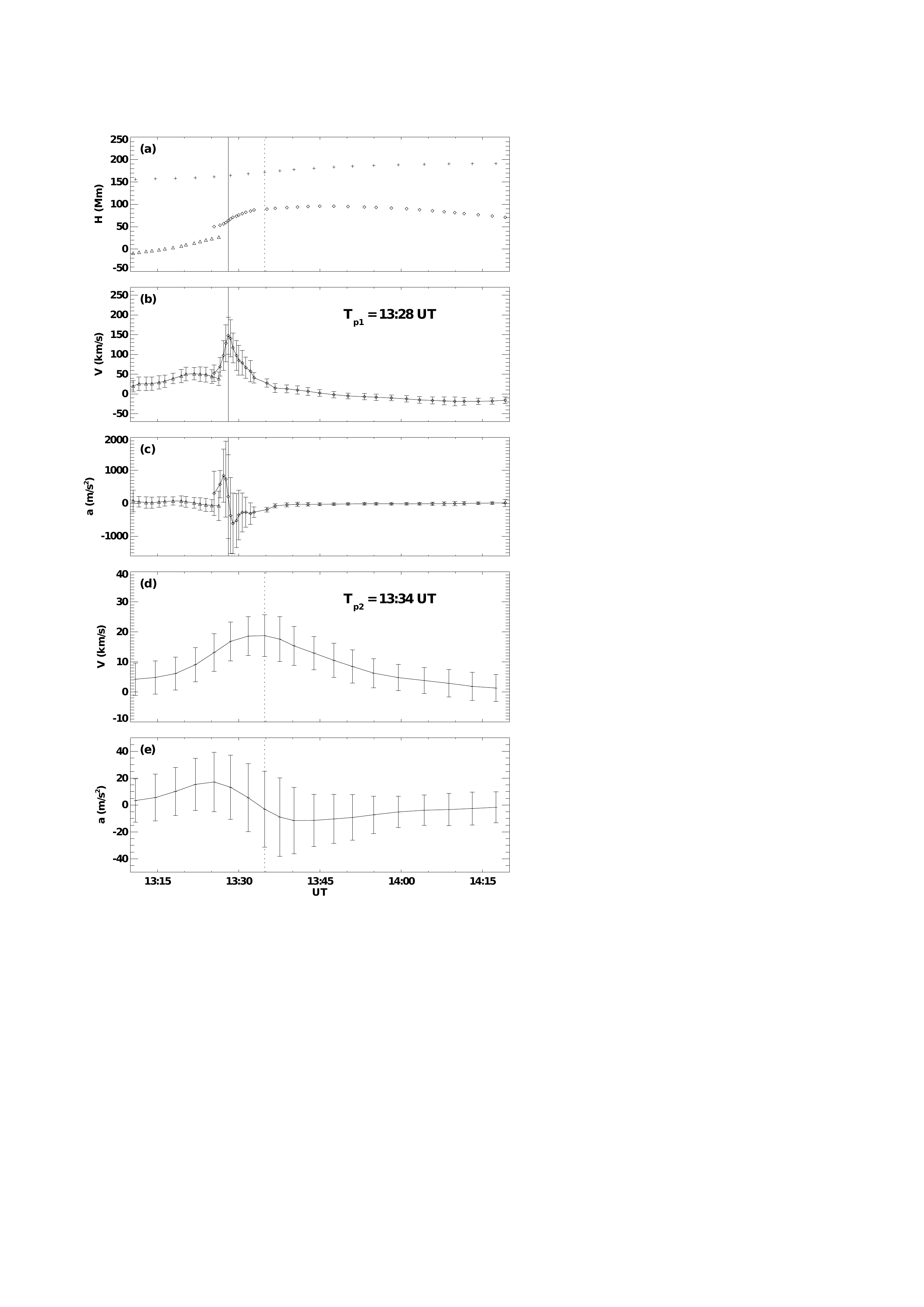}
               }

\vspace{0.0\textwidth}   
\caption{Dynamics of the up-boundaries marked in Figure~\ref{f6} (the triangles for the PFLs; the diamonds for the hot bright structure; the pluses for the overlying loop arcades). (a) for heights; (b) and (d) for velocities; (c) and (e) for accelerations.}
\label{f7}

\end{figure}

Based on the height-time plots of the leading edges (Figure~\ref{f7}(a)), we further derived their ascending velocities and accelerations and plotted them with uncertainties in Figure~\ref{f7}(b) and (c). From these plots, we found that the hot structure appeared around 13:20 UT (long after the flare onset time, 12:43 UT) at the height about 50 Mm, while the PFLs reached its maximum height about 30 Mm. Then it accelerated to its maximum velocity around 150 km/s at the time around 13:28 UT (solid vertical line in Figure~\ref{f7}), which was almost the same as the GOES SXR peak time (13:29 UT) considering the  uncertainties. This result suggests that the drive source for the ascending be related with the flare reconnection.

The overlying loop arcade was obvious in the cool 171 {\AA} plot (Figure~\ref{f6}(c)).As the same as above, we identified its leading edge and plotted the height-time, variations of velocities and accelerations in Figure~\ref{f7}(a), (d) and (e), respectively. From these plots, we found that the overlying loop arcade accelerated with the rising of the PFLs and the hot structure, until it reached its maximum velocity around 18 km/s at the time 13:34 UT, which was also the time when the rising of the hot structure stopped. Hence, there was a possible interaction between the overlying loop arcade and the rising hot structure, or, to put it another way, the overlying loop arcade prevented the hot structure from further eruption. Such failed eruption is consistent with the fact that there was no CME associated with the flare.

\subsection{Multiple Loop Systems}

\begin{figure} 
     \vspace{-0.0\textwidth}    
     \centerline{\hspace*{0.02\textwidth}
               \includegraphics[width=0.88\textwidth,clip=1]{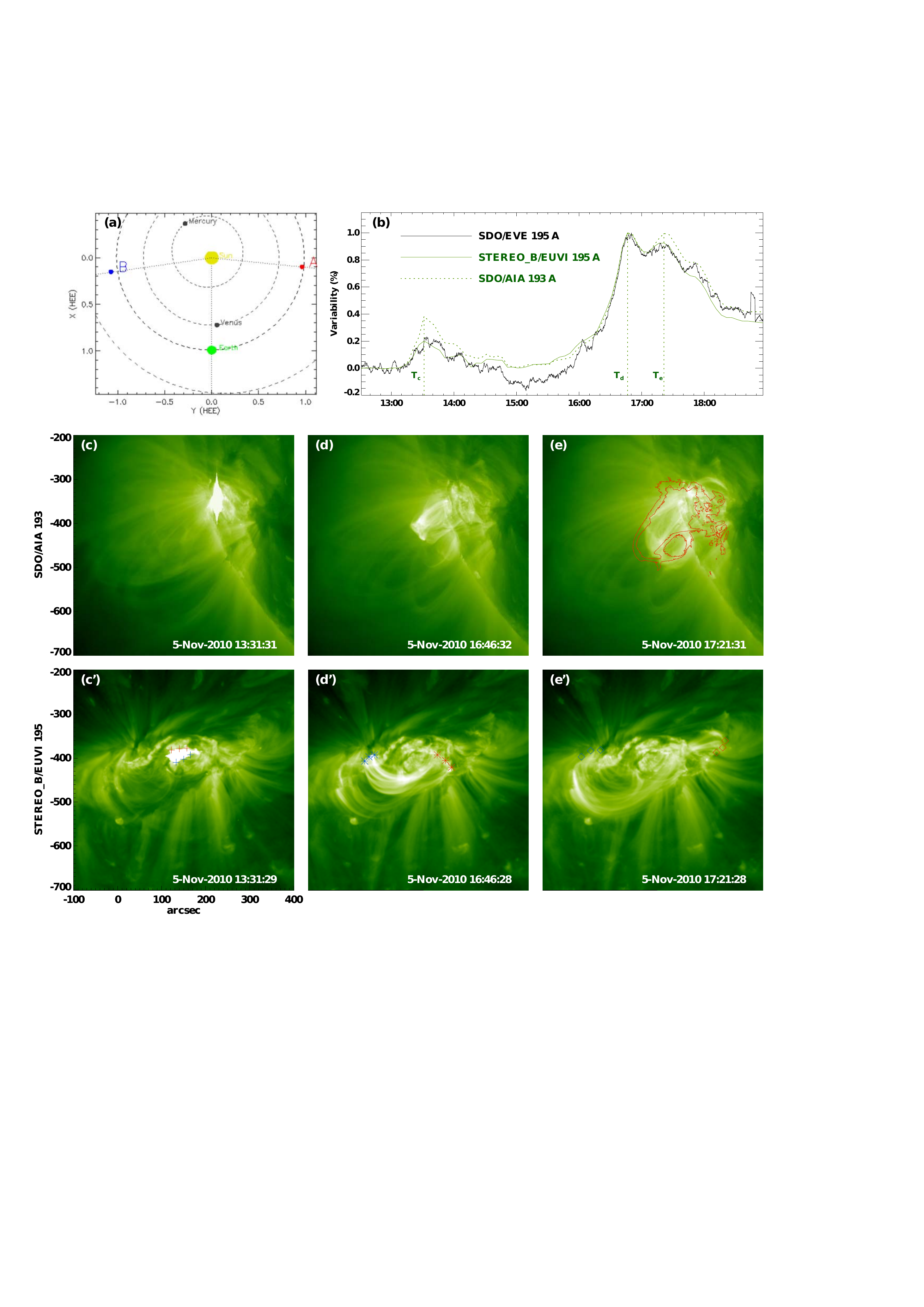}
               }

\vspace{0.0\textwidth}   
\caption{(a), Schematic for the positions of STEREO A and B on Nov. 5th, 2010. (b), Profiles of EVE 195 {\AA} line irradiance variability (black solid curve); region variability of STEREO\_B/EUVI 195 {\AA} (green solid curve) and SDO/AIA 193 {\AA} (green dotted curve). (c), (d) and (e) are the sub region snapshots of AIA 193 {\AA} images, corresponding to the three peak times T$_{c}$, T$_{d}$ and T$_{e}$ in panel (b), respectively; (c'), (d') and (e') are the sub region snapshots of STEREO\_B/EUVI 195 {\AA} images at the same peak times. The red curves in (e) are the same contours as in Figure~\ref{f4}(b). }
\label{f8}

\end{figure}

The \emph{Extreme UltraViolet Imager} (EUVI, \citealt{Wuelser2004}) on board  \emph{Solar TErrestrial Relations Observatory} (STEREO) also takes EUV images of the whole sun. Thanks to the unique positions of the STEREO A and B (see Figure~\ref{f8}(a)), the images from EUVI-B could provide new perspective of the flare event. Although EUVI has multiple passbands, we use the 195 {\AA} images, which havea cadence of 5 minutes and are comparable in temperature sensitivity with the EVE 19.5 nm and AIA 193 {\AA} observations.

During the flare time, AR 11121 located at the center of the EUVI-B 195 {\AA} images. As we did to AIA images, a sub-region including the AR was cut out from the images, then we summed the pixel values and plotted the region variability profile (greed solid curve in Figure~\ref{f8}(b)). Figure~\ref{f8}(b) also gave the profiles of EVE 195 {\AA} line irradiance variability (black solid curve) and AIA 193 {\AA} region variability (green dotted curve). The consistence of the three profiles indicates that the three different instruments were observing the same feature.

There were three obvious peaks in all the three profiles, which were the main phase peak (T$_{c}\sim$13:31 UT), the first (T$_{d}\sim$ 16:46 UT) and second (T$_{e}\sim$ 17:21 UT) late phase peak, respectively. By inspecting the image sequences of AIA 193 {\AA} and EUVI-B 195 {\AA} (\textbf{see the online STEREO animation}), we found that the three peaks corresponded to the brightening of three loop systems. For better illustration, the snapshots of both the AIA 193 {\AA} and EUVI-B 195 {\AA} sub images at three peak times were plotted in Figure~\ref{f8}(c)-(e) and \ref{f8}(c')-(e'). The loop system responsible for the first peak was the PFLs which were compact and south-north oriented (the red and blue pluses in Figure~\ref{f8}(c') represented their two footpoints, which were more obvious in the animation). The loop system for the second peak, or the late phase peak, was semicircular.. This loop arcade had a different orientation (east-west, Figure~\ref{f8}(d')) from that of the PFLs.It was more like the ``late phase loop arcade'' discussed by \citet{Liu2013}, since it had one footpoint (the red asterisks) near the PFLs, while another footpoint (blue asterisks) located at a remote site.

The third loop system was a sigmoid-like structure as seen in the EUVI-B images (Figure~\ref{f8}(e')). It had the same orientation (east-west; footpoints: the blue and red diamonds) as the second loop system, but it was wider and higher based on the Figure~\ref{f8}(d) and (e). The red curves in Figure~\ref{f8}(e) were the contours of the hot structure at its maximum height in AIA 131 {\AA} image (Figure~\ref{f4}(b)). The fact that the red curves could wrap the loop system very well suggests that the loop system might be the cool remnant of the hot structure. Thus, the late phase loop arcade and the remain of the failed-erupting hot structure together were responsible for the extremely large EUV late phase.

\section{Discussions}

The facts that i) the flare had a long decay phase even though a confined flare, ii) the flare had an extremely large late phase and iii) both the EVE 17.1 nm and AIA 171 {\AA} profiles had a significant dimming signature, suggested that there was an additional heating process during the gradual phase of this flare. To further confirm this, we applied the \emph{Deferential Emission Measure} (DEM; \citealt{Schmelz2011,Cheng2012}) method to the six AIA coronal passband images to create temperature maps (see Figure~\ref{f4}e and \textbf{the online Temperature animation}; \citealt{Song2014}). The same as for AIA images, we cut the slices from these maps and stacked as slice-time plots (Figure~\ref{f6}(d) and (h)). It was interesting to find that there was a jump of temperature in both plots around the time 14:30 UT. Assuming a single cooling or heating process, there would not be a jump in the temperature evolution. Therefore, the additional heating existed, and contributed to the extremely large EUV late phase. The beginning of this heating process was hard to diagnose, but the end time was probably at the valley of the dimming, $\sim$15:36 UT, and after that, it was the cooling process that dominated.

The heating source during the gradual phase was most likely originating from the failed-erupting hot structure. A similar failed eruption was studied by \citep{Song2014}, who identified that the erupting hot structure should be a magnetic flux rope. In our case, although lacking of direct magnetic observations, we found that the hot structure cooled down into lower temperature passband 195 {\AA} as a twisted loop system, which allowed us to consider that the hot structure was also a flux rope. This flux rope might have existed long before the eruption in the flare region, since it had a 20 Mm distance from the PFLs when it appeared. The leading edge velocity of the flux rope reached its maximum at the SXR peak time, indicating that the flux rope was energized by the flare reconnection. Following the end of the flare reconnection, the flux rope eruption motion was prevented by the strong overlying field. The interaction between the helical flux rope and the ambient coronal magnetic field may lead to slow magnetic reconnection, supplying additional energy source for flare heating in the decay phase \citep{Liu2014}. The indirect evidence was the temperature jump in Figure~\ref{f6}(d) started from top to middle, suggesting the heating source was at the top boundary.

\begin{figure} 
     \vspace{-0.0\textwidth}    
     \centerline{\hspace*{0.02\textwidth}
               \includegraphics[width=0.88\textwidth,clip=1]{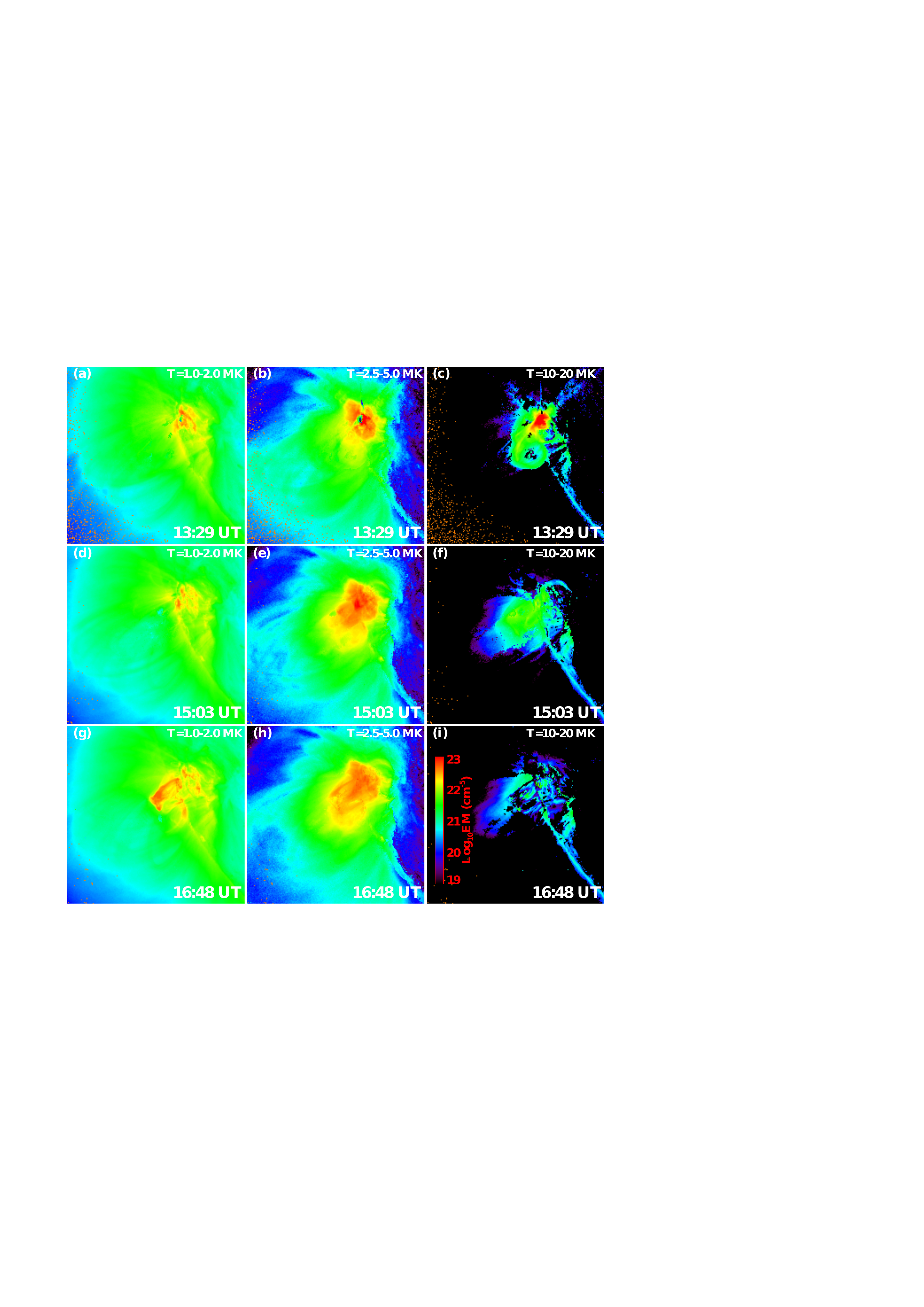}
               }

\vspace{0.0\textwidth}   
\caption{Emission maps from DEM analysis. The vertical columns from left to right are in three different temperature bands, 1.0 - 2.0 MK, 2.5 - 5.0 MK and 10 -20 MK. The rows from top to bottom are at three different times, 13:29 UT, 15:03 UT and 16:48 UT, corresponding to the three vertical lines in Figure~\ref{f3}, from left to right, respectively.}
\label{f9}

\end{figure}

\citet{Dai2013} also found multiple peaks in AIA late phase region profiles and used them as the evidence for the additional heating in their event. However, in our case, the multiple peaks were corresponding to different loop systems. One might find that it is hard to distinguish the late phase loop arcades from the twisted loop system, because the nature of low density of the flux rope made it obscure in the EUVI-B 193 {\AA} images. However, as a result of the DEM analysis, we plotted the emission maps in three different temperature bands at three different times as seen in Figure~\ref{f9}. Figure~\ref{f9} indicated that, the hottest emissions was originated from the PFLs during the main phase, while the warm emissions existed during all the flare phases. The cool emissions began to rise at the late phase, and the brightening features in Figure~\ref{f9}(g) and (h) were similar to Figure~\ref{f8}(d) and (e), thus they were corresponding to the two loop systems. The two loop systems had not only different sizes but also different thermal processes. The reason for only one late phase peak in the warmer temperature profiles (see Figure~\ref{f5}(c)) was probably due to the additional heating process, which maintained the two loop system had the same temperature before they cooled down to lower temperature.

\citet{Woods2014} pointed out that late phase flares had a duel-decay behavior in X-ray profiles, but this behavior was a poor proxy ($\sim$ 50\%) for EUV late phase flares. The event in this paper had a single decay slope as shown in X-ray flare, but had a EUV late phase. Moreover, as a preliminary statistic result, the LDE flare events with extremely large EUV late phase were mostly confined events. Although this result needs to be further confirmed and quantified through the investigation of a larger number of events, it might be a criterion for confined or eruptive flares.

\section{Conclusions}

We studied in detail a confined but LDE flare event with extremely large EUV late phase. To reconcile these observations, we found that the flare emissions were originated from multiple loop systems. The main phase was from a relatively small loop system forming the usual post-flare loop system, while the late phase possibly came from both a late phase loop arcade and a failed-erupting magnetic flux rope. The failed-erupting flux rope became the heating source after the main phase, possibly through a slow magnetic reconnection. The additional heating provided by the slow magnetic reconnection in gradual phase along with the long cooling process of the late phase loop arcade created the extremely large EUV late phase.

It is well known that the solar EUV radiation is a major energy source for creating and driving the Earth's ionosphere (\citealt{Liu2011,Woods2014}). The event we studied here had much more EUV radiation released during the late phase than that in the main phase. Thus, a statistic study of this kind of events will not only help to further understand the flare energy distribution in corona, but also study the effects of EUV late phase on the ionosphere, that will be the future work.


\acknowledgements
We thank the anonymous referee for suggestions that have improved this paper. CHIANTI is a collaborative project involving George Mason University, the University of Michigan (USA) and the University of Cambridge (UK). SDO is a mission of NASA's Living With a Star Program. We are grateful to the \emph{SDO}, \emph{STEREO} and \emph{SOHO} consortium for the free access to the data. K. Liu acknowledges the NSFC 41404134 and China Postdoctoral Science Foundation Funded Project (Project No. 2014M551814). This work was also supported by grants from NSFC  41131065, 41121003, 41304145; 973 key project 2011CB811403 and CAS Key Research Program KZZD-EW-01-4.

.

\bibliographystyle{apj}


\begin{table}
\setlength{\tabcolsep}{4pt}
\begin{center}
\caption{The properties of lines in Figure~\ref{f2}}
\label{t1}
\begin{tabular*}{0.68\textwidth}{l | l | c | c | c | c}
\hline\hline
\multirow{2}{*}{Ion} & $\lambda$ & $T_{max}$ & Main Phase & Dimming & Late Phase \\
\cline{2-6}
  & (nm) & $(log_{10})$ & \multicolumn{3}{|c }{$\bigtriangleup I\ (10^{-6}W/m^{2}/nm$)} \\
\hline
Fe XVIII & 9.4 & 6.9 & 28 & 24 & -  \\
Fe XIX & 10.8 & 7.0 & 30 & - & - \\
Fe XXI & 11.8 & 7.1 & 32 & - & - \\
Fe XX & 12.2 & 7.1 & 22 & - & - \\
Fe XXI & 12.9 & 7.1 & 42 & - & - \\
Fe XX & 13.3 & 7.1 & 84 & - & - \\
Fe XXII & 13.6 & 7.1 & 24 & - & - \\
Fe IX & 17.1 & 5.9 & -29 &  -40 & - \\
Fe X & 17.5 & 6.1 & - & - & 25 \\
Fe XI & 18.0 & 6.2 & - & - & 24 \\
Fe XII & 19.5 & 6.2 & - & - & 22 \\
Fe XIII & 20.3 & 6.3 & - & - & 21 \\
Fe XIV & 21.1 & 6.3 & - & - & 20 \\
He II & 25.6 & 4.9 & - & - & 25 \\
Fe XV & 28.4 & 6.4 & - & - & 73 \\
He II & 30.4 & 4.9 & 96 &  111 & 170 \\
Fe XVI & 33.5 & 6.8 & 43 & 59 & 68 \\
Fe XVI & 36.1 & 6.8 & - & 20 & 36 \\
Mg IX & 36.8 & 6.0 & - & - & 24 \\
\hline
\end{tabular*}
\end{center}
\end{table}


\end{document}